\documentclass[runningheads,orivec,envcountsame]{llncs}

\makeatletter \RequirePackage[bookmarks,unicode,colorlinks=true]{hyperref}%
\def\@citecolor{blue}%
\def\@urlcolor{blue}%
\def\@linkcolor{blue}%

\def\orcidID#1{\href{http://orcid.org/#1}{\smash{\protect\raisebox{-1.25pt}{\protect\includegraphics{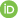}}}}}
\makeatother

\usepackage{amsmath, amssymb, amsfonts}
\usepackage{xcolor}
\usepackage{cite}
\usepackage{enumerate}
\usepackage{graphicx}
\usepackage{epstopdf}
\usepackage{bm}
\usepackage{todonotes}
\usepackage[noline]{algorithm2e}
\usepackage{stmaryrd}
\usepackage{hyperref}
\hypersetup{
    colorlinks,
    linkcolor={red!80!black},
    citecolor={green!60!black},
    urlcolor={blue}
}
\usepackage[capitalize]{cleveref}
\usepackage{mathtools}

\everypar{\looseness=-1}
\crefname{equation}{}{}
\crefname{lemma}{Lemma}{Lemmas}
\crefname{section}{sect.}{sect.}%
\crefname{figure}{fig.}{figures}%
\crefname{table}{table}{tables}%
\crefname{theorem}{thm.}{theorems}%
\crefname{corollary}{cor.}{corollaries}%
\crefname{proposition}{proposition}{propositions}%
\crefname{definition}{def.}{definitions}%
\crefname{example}{ex.}{ex.}%
\crefname{appendix}{App.}{App.}%

\newcommand{\IN}{\mathbb{N}}

\renewcommand{\epsilon}{\varepsilon}

\newcommand{\twodots}{\mathinner {\ldotp \ldotp}}

\newcommand{\RR}{\mathbb{R}}
\newcommand{\QQ}{\mathbb{Q}}

\newcommand{\ZZ}{\mathbb{Z}}
\newcommand{\NN}{\mathbb{N}}
\newcommand{\RA}{\mathbb{R}_{\mathbb{A}}}
\newcommand{\RAP}{\mathbb{R}_{\mathbb{A}>0}}
\newcommand{\RAPP}{\mathbb{R}_{\mathbb{A}\geq 0}}

\renewcommand{\dim}{d}

\newcommand{\up}{\mathrm{up}}

\newcommand{\wloop}[2]{\textbf{while} \quad #1 \quad \textbf{do} \quad #2 \quad \textbf{end}}

\newenvironment{myproof}{
  \noindent{\it Proof.}
}{\qed
  \medskip
}

\newcommand{\Def}{\mathrel{\mathop:}=}
\renewcommand{\phi}{\varphi}

\newcommand{\sign}{\mathsf{sign}}
\newcommand{\esign}{\mathsf{esign}}
\newcommand{\rootbound}{\mathsf{rootbound}}
\newcommand{\roots}{\mathsf{roots}}
\newcommand{\degree}{\mathsf{degree}}

\newcommand{\vecx}{\vec{x}}

\newcommand{\vecb}{\vec{b}}

\newcommand{\diag}{\mathrm{diag}}

\newcommand{\rb}{\mathit{rb}}
\newcommand{\ord}{\mathit{ord}}

\newcommand{\MRF}{\text{M}\Phi\text{RF}}

\newcommand{\mat}[1]{\left(\begin{smallmatrix} #1 \end{smallmatrix}\right)}

\title{On Deciding Constant Runtime of Linear Loops}

\author{Florian Frohn\inst{1}\orcidID{0000-0003-0902-1994} \and Jürgen Giesl\inst{1}\orcidID{0000-0003-0283-8520} \and Peter Giesl\inst{2}\orcidID{0000-0003-1421-6980} \and Nils Lommen\inst{1}\orcidID{0000-0003-3187-9217}}
\institute{RWTH Aachen University,  Aachen, Germany \and Department of Mathematics, University of Sussex, Brighton, UK}

\allowdisplaybreaks

\begin{document}

\maketitle

\begin{abstract}
  We consider linear single-path loops of the form
  \[
    \wloop{\varphi}{
      \vecx \gets A \vecx + \vecb
    }
  \]
  where $\vecx$ is a vector of variables, the loop guard $\varphi$ is a conjunction of linear inequations over the variables $\vecx$, and the update of the loop is represented by the matrix $A$ and the vector $\vecb$.
  It is already known that termination of such loops is decidable.
  In this work, we consider loops where $A$ has real eigenvalues, and prove that it is decidable whether the loop's runtime (for all inputs) is bounded by a constant if the variables range over $\RR$ or $\QQ$.
  This is an important problem in automatic program verification, since safety of linear \textbf{while}-programs is decidable if all loops have constant runtime, and it is closely connected to the existence of multiphase-linear ranking functions, which are often used for termination and complexity analysis.
  To evaluate its practical applicability, we also present an implementation of our decision procedure.
\end{abstract}

\section{Introduction}
\label{sect:introduction}

We investigate decidability of the question whether a linear\footnote{In this paper, the term ``linear'' refers to linear \emph{polynomials}, i.e., functions of the form $c_0 + \sum_{i=1}^d c_i \cdot x_i$ (in contrast to linear \emph{transformations} of the form $\sum_{i=1}^d c_i \cdot x_i$).\label{fn:affine}
}
single-path loop has constant runtime (i.e., whether the runtime of the loop can be bounded by the same constant for all inputs).
\begin{example}[Leading Example]
  \label{ex:leading}
  We consider linear single-path loops like
  \[
    \wloop{0 \leq x + y \leq 10}{
      \underbrace{\mat{
          x \\
          y
        }}_{\vecx}
      \gets
      \underbrace{\mat{
          1 & 0 \\
          0 & 2
        }}_{A}
      \mat{
        x \\y
      }
      +
      \underbrace{\mat{
          1 \\
          0
        }}_{\vecb}
    }
  \]

  \vspace*{-.1cm}

  \noindent
  with linear arithmetic only, conjunctive loop guards, and without branching in the loop body.
  In our example, the loop body is evaluated if $0 \leq x + y \leq 10$.
  The value of $x$ is increased by $1$ and $y$ is doubled in every iteration.
  Our goal is to decide whether the runtime of such loops is bounded by a constant.
  Later we will show that the loop in our example has indeed constant runtime.
\end{example}

Throughout this paper, the coefficients that occur in the loop are always rational numbers.\footnote{Our approach also works if the coefficients are algebraic real numbers and the variables range over $\RR$.
  We restrict ourselves to rational coefficients for simplicity.}
Regarding the domain of the variables $\vec{x}$, we focus on loops over $\RR$, but our results also apply to the case that the variables range over $\QQ$.

The motivation for our work is three-fold.
First, it is obviously relevant in the context of automated complexity analysis, in order to prove constant runtime.
Second, the question whether a linear loop admits a \emph{multiphase-linear ranking function}
($\MRF$) can be reduced to the question whether another linear loop has constant runtime w.r.t.\ a given set of initial states \cite[Sect.\ 4.2]{Ben-AmramGenaim19}.
Thus, our work is an important step towards a solution for the long-standing open problem whether the existence of $\MRF$s is decidable for linear loops.
$\MRF$s are relevant for both automated termination and complexity analysis, since the existence of a $\MRF$ does not only prove termination, but it also proves that the runtime of the loop is at most linear \cite[Thm.\ 7]{Ben-AmramGenaim17}.
However, our approach does not consider initial conditions, so it cannot be used for the inference of $\MRF$s yet.
We intend to refine it to support initial conditions in future work.
Third, our work is of interest for all infinite-state verification techniques that use approximations to deal with loops, like abstract interpretation, symbolic execution, recurrence analysis, loop summarization, etc.
All of these techniques make use of approximations in the presence of loops to prevent divergence of the verification process, but at the price of losing precision.
Instead, a loop with constant runtime $c$ can simply be unrolled $c$ times.
Thus, reachability (and hence, safety) is decidable for all those linear \textbf{while}-programs where all loops have constant runtime.

The constant runtime problem turns out to be surprisingly challenging:
Termination of linear loops is decidable over the reals \cite{Tiwari04}, the rationals \cite{Braverman06}, and the integers \cite{Hosseini19}.
But even for classes of loops where decidability of termination has been known for decades~\cite{Tiwari04}, decidability of the constant runtime problem is far from obvious.
Intuitively, the reason is that all decidability results for termination of linear loops exploit \emph{eventual monotonicity}:
For loops with non-negative real eigenvalues, the variable values start to behave monotonically after finitely many loop iterations (where the loop guard is ignored), and negative eigenvalues can easily be eliminated.
For the constant runtime problem, this kind of reasoning is not suitable, as one cannot disregard finitely many loop iterations.

\paragraph{Contributions:}
After introducing our notion of loops formally in \Cref{sec:Preliminaries}, in \Cref{sec:mon}
we prove decidability of the constant runtime problem for linear loops over $\RR$ with real eigenvalues.
Our decision procedure encodes the constant runtime problem as a validity problem, and then exploits the restriction to real eigenvalues to remove one quantifier alternation, so that the resulting formula is amenable to Fourier-Motzkin variable elimination.
Our results immediately extend to loops over $\QQ$, but not over $\ZZ$, due to the use of Fourier-Motzkin variable elimination.
For loops over $\ZZ$, we show decidability of the constant runtime problem for linear loops with eigenvalues from $\{-1,0,1\}$ in \Cref{sec:integers}.
In \Cref{sec:related_work}, we discuss related work, and in \Cref{sec:conclusion} we evaluate our implementation and conclude.

\section{Preliminaries}
\label{sec:Preliminaries}

We consider loops of the form
\begin{equation}
  \label{loop:inhomogeneous}
  \wloop{\phi}{\vecx\gets A\vecx + \vecb} \tag{\sc Loop}
\end{equation}
where $\vec{x}$ is a vector of $d \geq 1$ variables, $\phi$ is a conjunction of linear inequations over $\vec{x}$ of the form $t \sim 0$ with ${\sim} \in \{{>},{\geq}\}$, $A\in\QQ^{\dim\times\dim}$, and $\vecb\in\QQ^\dim$.
We sometimes regard a conjunction of formulas $\phi \equiv \phi_1 \land \ldots \land \phi_m$ as a set $\{ \phi_1, \ldots, \phi_m \}$ such that we can write $\phi_i \in \phi$ for every conjunct $\phi_i$.
For any arithmetic expression $t$ containing variables from $\vec{x}$, we define the \emph{update function} $\up$ of \eqref{loop:inhomogeneous} as $\up(t) \Def t[\vec{x}/A\vec{x} + \vec{b}]$, where $[\vec{x}/A\vec{x} + \vec{b}]$ denotes the substitution which replaces each variable from $\vec{x}$ by the corresponding polynomial from $A\vec{x} + \vec{b}$.

The loop \eqref{loop:inhomogeneous} has \emph{constant runtime} if the following holds:
\begin{equation}
  \label{eq:constant}
  \tag{\sc Const}
  \exists c \in \NN.\ \forall \vec{x} \in \RR^d.\ \exists i \in \NN_{\leq c}.\ \neg\phi[\vec{x}/\up^i(\vec{x})]
\end{equation}
Here, $\up^i$ denotes the $i$-fold application of $\up$ and $\NN_{\leq c} = \{ n \in \NN \mid n \leq c \}$ for any $c \in \NN$.
Intuitively, \eqref{eq:constant} states that $c \in \NN$ is a bound on the runtime of \eqref{loop:inhomogeneous} if, for every possible initial value $\vec{x} \in \RR^d$, the loop guard is violated after $i \leq c$ iterations, i.e., $\neg\phi[\vec{x}/\up^i(\vec{x})]$ holds.

As in \cite{TriangularCAV2019}, we can restrict ourselves to \emph{non-negative loops} (where $A$ does not have negative eigenvalues) without loss of generality.
The reason is that \eqref{loop:inhomogeneous} has constant runtime if and only if
\begin{equation*}
  \wloop{\phi \land \phi[\vec{x} / A\vec{x} + \vec{b}]}{\vec{x} \gets A(A\vec{x} + \vec{b}) + \vec{b}}
\end{equation*}
or, equivalently,
\begin{equation}
  \label{loop:nonneg}
  \tag{\sc Non-Neg}
  \wloop{\phi \land \phi[\vec{x} / A\vec{x} + \vec{b}]}{\vec{x} \gets A^2\vec{x} + A\vec{b} + \vec{b}}
\end{equation}
has constant runtime.
Here, the loop \eqref{loop:nonneg} is obtained by \emph{chaining} the loop \eqref{loop:inhomogeneous}, i.e., each iteration of \eqref{loop:nonneg} corresponds to two subsequent iterations of \eqref{loop:inhomogeneous}.
Clearly, if $A$ only has real eigenvalues, then all eigenvalues of $A^2$ are non-negative.
Now, \Cref{lem:simp} is a direct consequence of \cite[Lemma 18]{LPAR2020}.
\begin{corollary}[Transformation to Non-Negative Loops]
  \label{lem:simp}
  \Cref{loop:inhomogeneous} has constant runtime iff \Cref{loop:nonneg} has constant runtime.
\end{corollary}

\section{Loops with Real Eigenvalues}
\label{sec:mon}

From now on, we only consider loops with real eigenvalues.
Then we can restrict ourselves to loops like \eqref{loop:inhomogeneous} where all eigenvalues of $A$ are non-negative without loss of generality due to \Cref{lem:simp}.
The advantage of loops like \eqref{loop:inhomogeneous} is that they admit suitable closed forms capturing the behavior of $n$ loop iterations by a single expression for each variable.
These closed forms are heavily used in complete techniques for termination analysis, invariant generation, and complexity analysis (see, e.g., \cite{Tiwari04,Braverman06,Ouaknine15,Hosseini19,LPAR2020,POPL2019,kovacs2010CompleteInvariantGeneration,TriangularCAV2019,FMSD2023}).

The remainder of this section is structured as follows:
First, we introduce closed forms for linear loops with real eigenvalues in \Cref{sect:Closed Forms}.
Then, we use these closed forms to reduce the constant runtime problem to a validity problem in \Cref{sec:From Constant Runtime to Validity}, i.e., from that point onward, our goal is to decide validity of a first-order formula \eqref{eq:closed} with a specific structure.
The arithmetic expressions in this formula contain the variables $\vec{x}$ from the loop under consideration, and an additional, specific variable that serves as a ``loop counter''.
Next, in \Cref{sec:roots} we show that the number of real roots of these arithmetic expressions w.r.t.\ the loop counter can be bounded by a constant that does not depend on the values of $\vec{x}$.
In \Cref{sec:qelim}, we exploit this fact to eliminate the innermost quantifier from \eqref{eq:closed}.
For this step, it is crucial that all eigenvalues are non-negative real numbers.
Without the innermost quantifier, the resulting formula is amenable to a variant of Fourier-Motzkin variable elimination that allows us to eliminate the variables $\vec{x}$, see \Cref{sec:esign,sec:elim}.
In this way, we obtain a univariate formula that contains both polynomial and exponential arithmetic.
Finally, we show how to decide validity of such formulas in \Cref{sec:decide}.

\subsection{Closed Forms}
\label{sect:Closed Forms}

Let $\RA$ denote the algebraic reals (i.e., $\RA$ contains every real number that is a root of a univariate, non-zero polynomial with integer coefficients).
Moreover, $\RAP$ denotes the positive algebraic reals.
Similar to, e.g., \cite[Thm.~3.1]{POPL2019} and \cite[Def.~3.4]{FMSD2023}, for loops with non-negative real eigenvalues one can compute \emph{poly-exponential} closed forms for the $n$-fold update of the loop.
They have the form
\begin{equation}
  \label{eq:polyexp}
  \tag{\sc PE}
  \sum_{i=1}^M a_i(\vec{x}) \cdot b_i^n \cdot n^{e_i}
\end{equation}
where $b_i \in \RAP$, $e_i \in \NN$, and $a_i \in \RA[\vec{x}]$ is a linear polynomial over the variables $\vec{x}$.
Here, we identify $a_i$ and the function $\vec{x} \mapsto a_i$ where we sometimes write $a_i(\vec{x})$ to make the variables $\vec{x}$ explicit, and we use the same convention for other (vectors of) expressions.
Note that the $b_i$ are the eigenvalues of the update matrix $A$.
We highlight that the constants in \eqref{eq:polyexp} are algebraic, as it is unclear how to compute with non-algebraic numbers, and thus closed forms involving such numbers would not be suitable for our decision procedure.

To compute such a closed form, the idea is to consider the Jordan normal form $J = \diag(B_1,\dots,B_N)$ of the update matrix $A$, where $A = S\cdot J\cdot S^{-1}$ for a suitable change-of-basis matrix $S\in\RA^{d\times d}$.
Here, each Jordan block $B$ is associated to an eigenvalue $b\in\RAPP$.
Note that the Jordan normal form (and all relevant computations regarding algebraic numbers) can always be computed in polynomial time \cite{cai1994ComputingJordanNormal}.
Then the $n$-fold update is represented by the matrix $A^n = (S\cdot J\cdot S^{-1})^n = S\cdot J^n\cdot S^{-1}$ (i.e., $S$ and $S^{-1}$ always cancel out).
Thus, one only has to compute a closed form for each Jordan block of $J^n = \diag(B_1^n,\dots,B_N^n)$.
For every Jordan block $B \in \RAPP^{d_B\times d_B}$, this closed form is
\[
  B^n =
  \mat{
    \,b^n
    & \,\binom{n}{1}\cdot b^{n-1}
    & \,\cdots
    & \,\binom{n}{d_B-1}\cdot b^{n-(d_B-1)}\, \\[4pt]
    0
    & b^n
    & \cdots
    & \binom{n}{d_B-2}\cdot b^{n-(d_B-2)} \\[4pt]
    \vdots
    & \vdots
    & \ddots
    & \vdots \\[4pt]
    0
    & \cdots
    & 0
    & b^n
  }
  \qquad \text{for all $n\in\NN$}
\]
if $b \neq 0$, or $B^n = 0\in\RAPP^{d_B\times d_B}$ for all $n \geq d_B$ if $b = 0$.
Note that the binomial coefficients yield the polynomial factors $n^{e_i}$ of the
poly-exponential expression \eqref{eq:polyexp}.
Furthermore, in combination with the change-of-basis matrix $S\in\RA^{d\times d}$, the coefficients in the polynomials $a_i(\vec{x})$ are determined.

Hence, for any loop of the form \eqref{loop:inhomogeneous} with non-negative real eigenvalues only, there is a vector $\vec{cl}$ of $d$ poly-exponential expressions of the form \eqref{eq:polyexp} and a constant $n_0\in\NN$ such that
\[
  \up^n(\vec{x}) = \vec{cl} \qquad \text{for all $n \in \NN$ with $n \geq n_0$}.
\]
More precisely, $n_0$ is the dimension $d_B$ of the largest Jordan block $B$ which is associated to the eigenvalue $b = 0$.

\begin{example}[Closed Forms, \Cref{ex:leading} cont.]
  \label{ex:closed}
  For \Cref{ex:leading}, we have the closed form
  \[
    \vec{cl} =
    \mat{
      x + n \\
      2^n y
    }.
  \]
  To see why the constant $n_0$ is needed for closed forms, consider the loop
  \begin{equation*}
    \wloop{x \leq 0}{
      \mat{
        x \\ y
      } \gets
      \mat{
        0 & 1 \\
        0 & 0
      }
      \mat{
        x \\ y
      }
      +
      \mat{
        0 \\ 1
      }
    }
  \end{equation*}
  \noindent
  Here, we have $n_0 = 2$, as the update matrix is already in Jordan normal form, its only eigenvalue is $0$, and it has a single Jordan block of size $2$.
  Then we obtain the closed form $\vec{cl} =
    \mat{
      1 \\
      1
    }
  $.
  However, for $n \in \{0,1\}$ we have
  \[
    \vec{cl}[n/0] =
    \mat{
      1 \\
      1
    } \neq
    \mat{
      x \\
      y
    } = \up^0(\vec{x})
    \quad \text{and} \quad
    \vec{cl}[n/1] =
    \mat{
      1 \\
      1
    } \neq
    \mat{
      y \\
      1
    } = \up^1(\vec{x}).
  \]
\end{example}

\subsection{From Constant Runtime to Validity}
\label{sec:From Constant Runtime to Validity}
We now reduce the constant runtime problem to a validity problem.
As a first step, recall the characterization of constant loops:
\begin{equation*}
  \tag{\ref{eq:constant}}
  \exists c \in \NN.\ \forall \vec{x} \in \RR^d.\ \exists i \in \NN_{\leq c}.\ \neg\phi[\vec{x}/\up^i(\vec{x})]
\end{equation*}
By ``shifting'' $c$ by one, we obtain the equivalent formula
\begin{equation}
  \label{eq:closed0}
  \tag{$\textsc{Const}_<$}
  \exists c' \in \NN.\ \forall \vec{x} \in \RR^d.\ \exists i \in \NN_{< c'}.\ \neg\phi[\vec{x}/\up^i(\vec{x})]
\end{equation}
which is more suitable for our further analysis (in particular in \Cref{sec:roots}).
Thus, if \eqref{eq:closed0} holds, then $c'-1$ is a bound on the runtime of \Cref{loop:inhomogeneous}.
We now show that \eqref{eq:closed0} is equivalent to
\begin{equation*}
  \exists c \in \NN.\ \forall \vec{x} \in \RR^d.\ \exists i \in \NN_{< c}.\ \neg\phi[\vec{x}/\up^{n_0 + i}(\vec{x})]
\end{equation*}
where $n_0$ is defined as in \Cref{sect:Closed Forms}
This allows us to directly replace $\up^{n_0 + i}(\vec{x})$ by its closed form, since $i \geq 0$.
To this end, we define the \emph{instantiated loop guard}
\[
  \psi \Def \phi[\vec{x}/\vec{cl}].
\]
So we show that \eqref{eq:closed0} is equivalent to
\begin{equation}
  \label{eq:closed}
  \tag{$\textsc{Const}_{n_0}$}
  \exists c \in \NN.\ \forall \vec{x} \in \RR^d.\ \exists i \in \NN_{< c}.\ \neg\psi[n/n_0 + i],
\end{equation}
and if \eqref{eq:closed} holds, then $n_0 + c - 1$ is a bound on the runtime of \Cref{loop:inhomogeneous}.

\subsubsection{Proving Equivalence of \eqref{eq:closed0} and \eqref{eq:closed}:}
To show that \eqref{eq:closed0} and \eqref{eq:closed} are equivalent, we first prove that \eqref{eq:closed} implies \eqref{eq:closed0}.
To this end, assume that \eqref{eq:closed} holds, but \eqref{eq:closed0} does not.
Hence, there exists a $c\in \NN$ such that
\begin{equation}
  \tag{$\textsc{Const}^{in}_{n_0}$}
  \forall \vec{x} \in \RR^d.\ \exists i \in \NN_{< c}.\ \neg\psi[n/n_0 + i].\label{*}
\end{equation}
As \eqref{eq:closed0} does not hold, for $c' = n_0 + c$ there is an $\vec{x} \in \RR^\dim$ such that
\[
  \forall i \in \NN_{< n_0 + c}.\ \phi[\vec{x}/\up^i(\vec{x})],
\]
which contradicts \eqref{*}, as for every $i \geq 0$ we have $\neg\psi[n/n_0 + i] \equiv \neg\phi[\vec{x}/\up^{n_0 + i}(\vec{x})]$.

To see why \eqref{eq:closed0} implies \eqref{eq:closed}, assume that \eqref{eq:closed0} holds, but \eqref{eq:closed} does not.
Hence, there exists a $c'\in \NN$ such that
\begin{equation}
  \tag{$\textsc{Const}^{in}_<$}
  \forall \vec{y} \in \RR^d.\ \exists i \in \NN_{< c'}.\ \neg \phi[\vec{x}/\up^i(\vec{y})].\label{**}
\end{equation}
As \eqref{eq:closed} does not hold, for $c=c'$ there exists an $\vec{x} \in \RR^\dim$ such that
\begin{equation}
  \tag{$\dagger$}
  \forall i \in \NN_{< c'}.\ \psi[n / n_0+i].\label{***}
\end{equation}
Now we set $\vec{y}=\up^{n_0}(\vec{x})$ in \eqref{**}, i.e., there exists an $i\in \NN_{< c'}$ such that
$\neg \phi[\vec{x}/\up^i(\vec{y})]$, or equivalently, $\neg \phi[\vec{x}/\up^{n_0+i}(\vec{x})]$ holds.
This is a contradiction to \eqref{***}, as for every $i \geq 0$ we again have $\psi[n/n_0 + i] \equiv \phi[\vec{x}/\up^{n_0 + i}(\vec{x})]$.

While \eqref{eq:closed0} expresses that the loop guard must be violated within the first $c'$ iterations, \eqref{eq:closed} ``ignores'' the first $n_0$ iterations.
This allows us to use closed forms which are only valid for $n \geq n_0$.

\begin{example}[Constant Runtime to Validity, \Cref{ex:closed} cont.]
  \label{ex:validity}
  For \Cref{ex:leading}, it remains to decide validity of
  \begin{equation}
    \label{eq:validity1}
    \tag{\sc Val}
    \exists c \in \NN.\ \forall x,y \in \RR.\ \exists i \in \NN_{< c}.\ \neg \left( \, 0 \leq x
    + i + 2^iy \leq 10 \,\right).
  \end{equation}
  Moreover, if \eqref{eq:validity1} holds, then $c-1$ is a bound on the runtime of \Cref{ex:leading}.
\end{example}

\subsection{Bounded Number of Real Roots}
\label{sec:roots}
Recall that \eqref{eq:closed} states that there is a $c \in \IN$ such that for all initial values $\vec{x}$, there is an iteration $n$ with $n_0 \leq n < n_0 + c$ where the loop guard does not hold.
So here, we have to consider all instantiations of $n$ with natural numbers from $n_0$ up to $n_0 + c - 1$.
For deciding validity of \eqref{eq:closed}, the problem is that $c$ is not yet known.
Our aim is to replace \eqref{eq:closed} by
\begin{equation}
  \label{eq:no-conjunction}
  \tag{\sc QE}
  \exists m \in \NN.\ \forall \vec{x} \in \RR^\dim.\ \bigvee_{j=0}^\rb \neg\psi [n/n_0 + j \cdot m],
\end{equation}
where we only check $\rb +1$ instantiations of $n$.
Here, $\rb$ is a number that only depends on $\psi$, but not on the actual instantiation of $\vec{x}$.
Thus, our goal is to replace the existentially quantified formula $\exists i \in \NN_{< c}.\ \neg\psi[n/n_0 + i]$ which depends on the (unknown) value $c$ by the disjunction $\bigvee_{j=0}^\rb \neg\psi[n/n_0 + j \cdot m]$, where $\rb$ is known and fixed.
So to eliminate the inner existential quantifier in \eqref{eq:closed}, we show that it suffices to consider a fixed number of ``sample points'' instead of all values $i \in \NN_{<c}$.
Here, the (existentially quantified) variable $m$ is used as the distance between our ``sample points''.

To determine the number $\rb$, $n$ is instantiated by real instead of just natural numbers.
Then, the number $\rb$ is chosen in such a way that for every vector $\vec{v}$ of real numbers, the expressions in the inequations of the formula $\psi[\vec{x}/\vec{v}]$ (which only contains the variable $n$) have at most $\rb$ real roots.
Note that the number of real roots is unbounded without the restriction to non-negative real eigenvalues.

To define $\rb$, for any poly-exponential expression $t$
as in \eqref{eq:polyexp}, we first define its $\rootbound$.
Here, $\rootbound(t)$ only depends on the number $M$ of addends and the exponents $e_i$,
but $\rootbound(t)$ does not depend on $\vec{x}$.
The following lemma shows that for every value $\vec{v} \in \RR^d$, $\rootbound(t)$ is an upper bound on the number of real roots of the expression $t[\vec{x}/\vec{v}]$.

\begin{lemma}[Bounded Number of Real Roots]
  \label{lem:roots}
  Let $\vec{v} \in \RR^d$ and let
  \[
    t = \sum_{i=1}^M a_i(\vec{x}) \cdot b_i^n \cdot n^{e_i}
  \]
  be a poly-exponential expression such that $t[\vec{x}/\vec{v}]$ is not the polynomial $0$.
  Then
  \[
    \rootbound(t) \Def M - 1 + \sum_{i=1}^M e_i
  \]
  is an upper bound on the number of real roots of $t[\vec{x}/\vec{v}]$.
\end{lemma}
\begin{myproof}
  As in \cite{polya1998ProblemsTheoremsAnalysis}, for any expression $s$ of the form
  \begin{equation}
    \label{eq:ord}
    \sum_{i=1}^{M'} p_i(n) \cdot b_i^n,
  \end{equation}
  where each $b_i$ is unique and each $p_i(n)$ is a polynomial over $n$, we define
  \begin{equation}
    \label{eq:ord2}
    \ord(s) \Def \sum_{i=1}^{M'} (\degree(p_i(n)) + 1) = M' + \sum_{i=1}^{M'} \degree(p_i(n)).
  \end{equation}
  Let $\vec{v} \in \RR^\dim$ be arbitrary but fixed.
  Then we can rewrite
  \[
    t' = t[\vec{x}/\vec{v}] = \sum_{i=1}^M a_i(\vec{v}) \cdot b_i^n \cdot n^{e_i}
  \]
  to the form \eqref{eq:ord}, and thus by \cite[Part V, Chap.\ 1,
    Num.\ 75]{polya1998ProblemsTheoremsAnalysis}, it has at most $\ord(t') - 1$ real roots
  (where $t'$ contains no other variable than $n$).
  As $M' \leq M$ and $\sum_{i=1}^{M'}\degree(p_i(n)) \leq \sum_{i=1}^M e_i$, this
  implies that $t'$ has at most $M - 1 + \sum_{i=1}^M e_i$ real roots due to \eqref{eq:ord2}.
  As this bound is independent from $\vec{v}$, the claim follows.
\end{myproof}

We can now define the desired number $\rb$.

\begin{definition}[Root Bound of Loop Guard]
  We define the \emph{root bound} of the instantiated loop guard $\psi$ as
  \[
    \rb \Def \sum_{{(t \sim 0)} \in \psi} \rootbound(t),
  \]
  where ${\sim} \in \{ >, \geq \}$, i.e., here we consider all inequations of the form $t >0$ and $t \geq 0$ occurring in the conjunction $\psi$.
\end{definition}

\begin{example}[Root Bound, \Cref{ex:validity} cont.]
  \label{ex:k}
  For \Cref{ex:leading}, $\psi = \phi[\vec{x}/\vec{cl}]$ is
  \[
    0 \leq x + n + 2^n y \leq 10 \quad \equiv \quad x + n + 2^ny \geq 0 \; \land \; 10 - x - n - 2^ny \geq 0.
  \]
  Here,
  we have
  \[
    \rootbound(x + n + 2^ny) = \rootbound(10 - x - n - 2^ny) = 3
  \]
  since $M = 3$ and $\sum_{i=1}^3 e_i = 1$ and thus, $\rb = 3 + 3 = 6$.
\end{example}

\subsection{Eliminating the Innermost Quantifier}
\label{sec:qelim}

Now we show that \eqref{eq:closed} is equivalent to
\begin{align*}
  \exists m \in \NN.\ \forall \vec{x} \in \RR^\dim.\ \bigvee_{j=0}^\rb \neg \psi[n/n_0 +
    j \cdot m],
  \tag{\ref{eq:no-conjunction}}
\end{align*}
and if $\eqref{eq:no-conjunction}$ holds, then $n_0 + \rb \cdot m$ is a bound on the runtime of \Cref{loop:inhomogeneous}, again provided that all eigenvalues are non-negative reals.

Recall that \eqref{eq:no-conjunction} states that there is a distance $m$ such that for all initial values of $\vec{x}$, the guard is violated after $n_0$, $n_0 + 1 \cdot m$, \ldots, or $n_0 + \rb \cdot m$ iterations.

\subsubsection{Proving Equivalence of \eqref{eq:closed} and \eqref{eq:no-conjunction}:}
To see why \eqref{eq:closed} and \eqref{eq:no-conjunction} are equivalent, note that
\begin{equation}
  \label{eq:inst1}
  \tag{$\textsc{QE} \Rightarrow \textsc{Const}_{n_0}$}
  \text{\eqref{eq:no-conjunction} implies \eqref{eq:closed} by choosing $c = \rb \cdot m + 1$.}
\end{equation}
Thus, $n_0 + c - 1 = n_0 + \rb \cdot m$ is a bound on the runtime of \Cref{loop:inhomogeneous}.

To show that \eqref{eq:closed} implies \eqref{eq:no-conjunction}, assume \eqref{eq:closed}, i.e., there is a $c\in\NN$ so that $\forall \vec{x} \in \RR^\dim.\ \exists i \in \NN_{< c}.\ \neg\psi[n/n_0 + i]$.
Then for every $j \in \NN$, we have:
\begin{align*}
              & \forall \vec{x} \in \RR^\dim.\ \exists i \in \NN_{< c}.\ \neg\psi[n/n_0 + i] \tag{for some $c \in \NN$ by \eqref{eq:closed}} \\
  \iff \      & \forall \vec{x} \in \RR^\dim.\ \exists i \in \NN_{<
  c}.\ \neg\phi[\vec{x}/\vec{cl}][n/n_0+i] \tag{def.\ of $\psi$} \\
  \iff \      & \forall \vec{x} \in \RR^\dim.\ \exists i \in \NN_{< c}.\ \neg\phi[\vec{x}/\vec{cl}[n/n_0+i]] \tag{as $n$ does not occur in $\phi$} \\
  \iff \      & \forall \vec{x} \in \RR^\dim.\ \exists i \in \NN_{< c}.\ \neg\phi[\vec{x}/\up^{n_0+i}(\vec{x})] \tag{def.\ of $\vec{cl}$, as $i \geq 0$} \\
  \implies \  & \forall \vec{x} \in \RR^\dim.\ \exists i \in \NN_{<
  c}.\ \neg\phi[\vec{x}/\up^{n_0+j \cdot c+i}(\vec{x})] \tag{for any $j \in \NN\ (\ddagger)$} \\
  \iff \      & \forall \vec{x} \in \RR^\dim.\ \exists i \in \NN_{< c}.\ \neg\phi[\vec{x}/\vec{cl}[n/n_0 + j \cdot c + i]] \tag{def.\ of $\vec{cl}$, as $j \cdot c + i \geq 0$} \\
  \iff \      & \forall \vec{x} \in \RR^\dim.\ \exists i \in \NN_{< c}.\ \neg\phi[\vec{x}/\vec{cl}][n/n_0 + j \cdot c + i] \tag{as $n$ does not occur in $\phi$} \\
  \iff \      & \forall \vec{x} \in \RR^\dim.\ \exists i \in \NN_{< c}.\ \neg\psi[n/n_0
    + j \cdot c + i] \tag{def.\ of $\psi$}
\end{align*}
For the step $(\ddagger)$, $\vec{x}$ is again universally quantified, i.e., we can instantiate $\vec{x}$ by $\up^{j \cdot c}(\vec{x})$, as in the proof that \eqref{eq:closed0} implies \eqref{eq:closed}, where we instantiated the universally quantified variables $\vec{y}$ by $\up^{n_0}(\vec{x})$.
Thus, \eqref{eq:closed} implies
\begin{equation}
  \label{eq:sign-changes}
  \tag{\sc InIntervals}
  \forall \vec{x} \in \RR^\dim.\ \bigwedge_{j=0}^\rb \exists i \in \NN_{< c}.\ \neg\psi[n/n_0 + j \cdot c + i].
\end{equation}
So \eqref{eq:sign-changes} holds iff for all initial values of $\vec{x}$ and all $0 \leq j \leq \rb$, the loop guard is violated
after $n_0 + j \cdot c$, $n_0 + j \cdot c + 1$, \ldots, or $n_0 + (j+1) \cdot c - 1$ iterations.

To prove that \eqref{eq:sign-changes} implies \eqref{eq:no-conjunction}, we assume that \eqref{eq:sign-changes} holds and \eqref{eq:no-conjunction} does not hold, and derive a contradiction.
If \eqref{eq:no-conjunction} does not hold, then there is a $\vec{v} \in \RR^\dim$ such that
\begin{equation}\label{eq:sign-changes-contra}
  \tag{\sc AtBorders}
  \bigwedge_{j=0}^\rb \psi[n/n_0 + j \cdot c] [ \vec{x} / \vec{v} ]
\end{equation}
by choosing $m = c$ in \eqref{eq:no-conjunction}.
Then by the intermediate value theorem, for each $0 \leq j \leq \rb$, there is at least one ${(t_j \sim_j 0)} \in \psi$ with ${\sim_j} \in \{ >, \geq \}$ such that $t_j [ \vec{x} / \vec{v} ]$ has a real root in the interval
\begin{equation}
  \label{interval}
  [n_0 + j \cdot c, \; n_0 + (j+1) \cdot c - 1]
\end{equation}
due to \eqref{eq:sign-changes}, if we consider instantiations of $n$ with real instead of just natural numbers.
The reason is that \eqref{eq:sign-changes-contra} implies $t[ \vec{x} / \vec{v} ][n/n_0 + j \cdot c] \sim 0$ for all ${(t \sim 0)} \in \psi$, and \eqref{eq:sign-changes} implies that there exists a value $k \in [n_0 + j \cdot c, \; n_0 + (j+1) \cdot c - 1]$ and a ${(t_j \sim_j 0)} \in \psi$ such that $t_j[ \vec{x} / \vec{v} ][n/k] \sim_j 0$ is violated.
Then we must also have $t_j[ \vec{x} / \vec{v} ][n/k_0] = 0 \neq t_j[ \vec{x} / \vec{v} ][n/k_{\neq 0}]$ for some $k_0,k_{\neq 0} \in [n_0 + j \cdot c, \; n_0 + (j+1) \cdot c - 1]$, since $t_j[ \vec{x} / \vec{v} ]$ is a continuous function in the variable $n$.\footnote{The reason for continuity of $t_j[ \vec{x} / \vec{v}
  ]$ is that all exponential expressions in $t_j[ \vec{x} / \vec{v}
  ]$ have the form $b_i^n$ for $b_i \in \RR_{\geq 0}$, since all eigenvalues of the loop are non-negative reals.}
In other words, for every interval \eqref{interval} there is a $t_j[ \vec{x} / \vec{v} ]$ that satisfies $t_j[ \vec{x} / \vec{v} ] \sim_j 0$ at the ``borders'' of the interval and violates $t_j[ \vec{x} / \vec{v} ] \sim_j 0$ inside the interval.
Thus, we have $t_j[ \vec{x} / \vec{v} ] \neq 0$ (i.e., $t_j[ \vec{x} / \vec{v} ]$ is not
the polynomial $0$), and it must have a real root in the interval.
Hence, the sum of the number of real roots of all non-zero expressions $t[ \vec{x} / \vec{v} ]$, $(t \sim 0) \in \psi$, is at least $\rb+1$, i.e.,
\[
  \sum_{\mathclap{\substack{{(t \sim 0)} \in \psi\\t[\vec{x}/\vec{v}] \neq 0}}} |\roots(t [\vec{x}/\vec{v}])| > \rb
  = \sum_{\mathclap{{(t \sim 0)} \in \psi}} \rootbound(t).\pagebreak[3]
  \]
This contradicts the fact that $\rootbound(t)$ is a bound on the number of real roots of $t [\vec{x} / \vec{v}]$ by \Cref{lem:roots}.
Thus, as we instantiated $m$ with $c$ to derive \eqref{eq:sign-changes-contra}, we showed
\[
  \text{$\neg\eqref{eq:no-conjunction}$ and \eqref{eq:closed} yield a contradiction by choosing $m=c$}
\]
or, equivalently,
\begin{equation}
  \label{eq:inst2}
  \tag{$\textsc{Const}_{n_0} \Rightarrow \textsc{QE}$}
  \text{\eqref{eq:closed} implies \eqref{eq:no-conjunction} by choosing $m=c$.}
\end{equation}

\begin{example}[Quantifier Elimination, \Cref{ex:k,ex:validity} cont.]
  \label{ex:no-conjunction}
  For \eqref{eq:validity1}, it remains to decide validity of
  \begin{equation}
    \label{eq:no-conjunction1}
    \tag{$\textsc{Val}^\lor$}
    \exists m \in \NN.\ \forall x,y \in \RR.\ \bigvee_{j=0}^6
    \neg \left( \, 0 \leq x + j \cdot m + 2^{j \cdot m}y \leq 10 \, \right).
  \end{equation}
  In our example we have $n_0 = 0$ and $\rb = 6$.
  Thus, if \eqref{eq:no-conjunction1} holds, then this yields the bound $n_0 + \rb \cdot m = 6 \cdot m$ on the runtime of the loop in \Cref{ex:leading}.
\end{example}

The proof above yields the following corollary, which will be needed later.
\begin{corollary}
  \label{cor:large-models}
  Consider the subformula
  \begin{equation}
    \label{eq:matrix}
    \tag{$\textsc{QE}^{in}$}
    \forall \vec{x} \in \RR^\dim.\ \bigvee_{j=0}^\rb \neg\psi[n/n_0 + j \cdot m]
  \end{equation}
  of \eqref{eq:no-conjunction}, whose only free variable $m$ ranges over $\NN$.
  Then \eqref{eq:matrix} is either unsatisfiable, or it is valid for infinitely many different values of $m$.
\end{corollary}
\begin{myproof}
  First assume $\rb = 0$.
  Then \eqref{eq:matrix} does not depend on $m$, and thus it is either unsatisfiable, or it is valid for \emph{all} possible values of $m$.

  Now let $\rb > 0$ and assume that $\eqref{eq:matrix}$ is only valid for finitely many values of $m$.
  Let $v \in \NN$ be the maximal number such that instantiating $m$ by $v$ satisfies $\eqref{eq:matrix}$.
  Then with \eqref{eq:inst1}, it follows that instantiating $c$ with $rb\cdot v+1$ satisfies \eqref{*}.
  Next, by \eqref{eq:inst2}, it follows that instantiating $m$ with $rb\cdot v+1$ satisfies \eqref{eq:matrix}.
  As $\rb > 0$, we have $\rb \cdot v + 1 > v$, i.e., \eqref{eq:matrix} also holds for a number larger than $v$, which contradicts our assumption on the maximality of $v$.
\end{myproof}

\subsection{Eventual Signs}
\label{sec:esign}
Compared to the formula \eqref{eq:closed}, in \eqref{eq:no-conjunction}
we eliminated the innermost quantifier and replaced it by $\rb +1$ subformulas, where $\rb$ is fixed.
The resulting formula \eqref{eq:no-conjunction} is linear in the program variables $\vec{x}$ and poly-exponential in $m$.

Next, we want to get rid of $\vec{x}$ by using a variant of Fourier-Motzkin elimination.
To see why Fourier-Motzkin elimination can be applied even though we have non-linear terms in $m$, we disregard the outermost quantifier of \eqref{eq:no-conjunction} for now, i.e., we assume that $m \in \NN$ is given.
Then proving \eqref{eq:no-conjunction} amounts to proving the\linebreak[3]
\emph{validity} of a \emph{disjunction} of inequations.
In contrast, Fourier-Motzkin elimination is used to check the \emph{satisfiability} of \emph{conjunctions}.
Thus, to prove validity of \eqref{eq:no-conjunction}, our goal is to use Fourier-Motzkin elimination to prove unsatisfiability of
\[
  \pi \Def \bigwedge_{j=0}^\rb \psi[n/n_0 + j \cdot m],
\]
which results from the negation of \eqref{eq:no-conjunction}.

However, Fourier-Motzkin's technique applies to linear inequations only.
To apply it in our setting, we regard the (poly-exponential) inequations in $\pi$ as linear inequations over $\vec{x}$ whose coefficients are univariate poly-exponential expressions w.r.t.\ $m$, i.e., the coefficients are of the form
\begin{equation}
  \label{eq:coeff}
  \tag{\ensuremath{\textsc{Coeff}}}
  0\qquad\text{or}\qquad\sum_{i=1}^M d_i \cdot b_i^m \cdot m^{e_i}
\end{equation}
where $m$ is the only variable, $M \geq 1$, $d_i \in \RA \setminus \{0\}$, $b_i \in \RAP$, and $e_i \in \NN$.
Again, note that $d_i$ and $b_i$ are algebraic numbers, and thus they are suitable for algorithmic purposes.
To apply Fourier-Motzkin elimination, the signs of these coefficients must be known, which is not the case, in general.

However, we can easily identify the asymptotically dominant addend of the expression \eqref{eq:coeff}, which determines its sign \emph{for large enough values of $m$}.
Moreover,
by \Cref{cor:large-models}, we may assume that $m$ is sufficiently large without loss of generality: If \eqref{eq:no-conjunction} holds for some small value of $m$, then there is also an arbitrarily large value of $m$ where \eqref{eq:no-conjunction} holds.

More formally, to determine the sign for sufficiently large $m$, w.l.o.g.\ assume that the pair $(b_i,e_i)$ is unique for each addend in \eqref{eq:coeff}.
Then it suffices to consider the \emph{asymptotically dominant} addend $d_j \cdot b_j^m \cdot m^{e_j}$ where $(b_j,e_j) \geq_{lex} (b_i,e_i)$ for all $1 \leq i \leq M$, i.e., $(b_j,e_j)$ is maximal w.r.t.\ the lexicographic ordering.
Since all eigenvalues $b_i$ are non-negative reals, for sufficiently large $m$, the sign of the coefficient \eqref{eq:coeff} is equal to the sign of $d_j$ or the coefficient is always $0$.
More precisely, for any number $v$, let:
\[
  \sign(v) = -1 \text{ if } v < 0 \qquad \sign(0) = 0 \qquad \sign(v) = 1 \text{ if } v > 0
\]
Note that since we required $d_i \neq 0$ and uniqueness of $(b_i,e_i)$ for each addend, when instantiating $m$ by large enough numbers, then we have $\sign(\ref{eq:coeff}) \neq 0$ whenever $\eqref{eq:coeff}$ is of the form $\sum_{i=1}^M d_i \cdot b_i^m \cdot m^{e_i}$.
In this case, we have
\[
  \exists m_0 \in \NN.\ \forall m \geq m_0.\ \sign(\ref{eq:coeff}) = \sign(d_j).
\]
We define the \emph{eventual sign} of \eqref{eq:coeff} as $\esign(\ref{eq:coeff}) \Def 0$ if \eqref{eq:coeff} is $0$ and as $\esign(\ref{eq:coeff}) \Def \sign(d_j)$ otherwise.
\begin{example}[Eventual Signs, \Cref{ex:no-conjunction} cont.]
  \label{ex:coeffs}
  Reconsider \eqref{eq:no-conjunction1}.
  Here, the coefficients for the variable $x$ are 1 and $-1$, and we obviously have $\esign(1) = 1$ and $\esign(-1) = -1$.
  The coefficients for the variable $y$ are $2^{j \cdot m}$ and $- 2^{j \cdot m}$ for $0 \leq j \leq 6$, where $\esign(2^{j \cdot m}) = 1$ and $\esign(-2^{j \cdot m}) = -1$.
  For a more complex example, consider $(3^m \cdot m^2 - 10 \cdot 2^m \cdot m^3) \cdot x$.
  As $(3,2) \geq_{lex} (2,3)$, we have $\esign(3^m \cdot m^2 - 10 \cdot 2^m \cdot m^3) = 1$, i.e., the asymptotically dominant coefficient of $x$ eventually becomes positive.
\end{example}

\subsection{Variable Elimination}
\label{sec:elim}
We now show how to eliminate a single variable $x$ in the vector $\vec{x}$ of variables from $\pi$.
Then it immediately follows that \emph{all} variables from $\vec{x}$ can be eliminated, so that the only remaining variable is $m$.
Recall that $\pi$ is defined as
\[
  \pi \Def \bigwedge_{j=0}^\rb \psi[n/n_0 + j \cdot m].
\]
So in particular, $\pi$ is a conjunction of inequations.
To eliminate $x$, consider the sets
\begin{align*}
  \check{\pi}_x & {} \Def \{p \cdot x + t \sim 0 \text{ occurs in } \pi \mid
  {\sim} \in \{{>},{\geq}\}, \esign(p) > 0\} \\
  \hat{\pi}_x   & {} \Def \{p \cdot x + t \sim 0 \text{ occurs in } \pi
  \mid
  {\sim} \in \{{>},{\geq}\}, \esign(p) < 0\}
\end{align*}
where $p$ is the coefficient of $x$ when regarding $p \cdot x + t$ as a linear polynomial w.r.t.\ $\vec{x}$ with coefficients of the form \eqref{eq:coeff}.
In other words,
$p$ is chosen such that $t$ does not contain any occurrence of $x$.
Thus, $\check{\pi}_x$ contains those inequations from $\pi$ that give rise to lower bounds on $x$ for large enough $m$, and $\hat{\pi}_x$ contains those inequations that give rise to upper bounds.
Next, we define $\pi_x$ to be the smallest set such that the following holds:
\[ \begin{array}{r@{\quad}l}
    \text{If}   & p_1 \cdot x + t_1 \sim_1 0 \in \check{\pi}_x \qquad\quad \text{and} \qquad\quad
    p_2 \cdot x + t_2 \sim_2 0 \in \hat{\pi}_x,                                                   \\
    \text{then} & p_1 \cdot t_2 - p_2 \cdot t_1 \sim 0 \in \pi_x,
  \end{array}
\]
where $\sim$ is $\geq$ if both ${\sim_1}$ and ${\sim_2}$ are ${\geq}$, and ${\sim}$ is ${>}$, otherwise.
Then
\[
  \pi \equiv \pi' \; \text{for all large enough values of $m$},
\]
where $\pi' \Def (\pi \cup \pi_x) \setminus (\check{\pi}_x \cup \hat{\pi}_x)$ and by
construction, $x$ does not occur in $\pi'$.
\begin{example}[Fourier-Motzkin Elimination, \Cref{ex:coeffs,ex:no-conjunction} cont.]
  \label{ex:elim}
  For the formula \eqref{eq:no-conjunction1}, we get the following inequations:
  \begin{align*}
    0                              & {} \leq \overbrace{1}^{p} \cdot\; x + \overbrace{j \cdot m + 2^{j \cdot m}y}^{t} \tag{for all $0 \leq j \leq 6$} \\
    x + j \cdot m + 2^{j \cdot m}y & {} \leq 10 \tag{for all $0 \leq j \leq 6$}
  \end{align*}
  Thus, we have:
  \begin{align*}
    \check{\pi}_x & {} = \{x + j \cdot m + 2^{j \cdot m}y \geq 0 \mid 0 \leq j \leq 6\} \\
    \hat{\pi}_x   & {} = \{-x - j \cdot m - 2^{j \cdot m}y + 10 \geq 0 \mid 0 \leq j \leq 6\}
  \end{align*}
  Hence, eliminating $x$ yields\pagebreak[3]
  \begin{align*}
    \pi' {} & = \{j \cdot m + 2^{j \cdot m}y \geq j' \cdot m + 2^{j' \cdot m}y - 10 \mid
    j, j' \in \{0,\ldots,6\} \} \\
            & = \{(j-j') \cdot m + (2^{j \cdot m} - 2^{j' \cdot m}) \cdot y + 10 \geq 0
    \mid j, j' \in \{0,\ldots,6\} \} \\
            & = \{(j-j') \cdot m + ((2^j)^{m} - (2^{j'})^m) \cdot y + 10 \geq 0 \mid
    j, j' \in \{0,\ldots,6\} \} \tag{$\star$} \label{eq:ub}
  \end{align*}
  Next, we eliminate $y$ from $\pi'$, where we get:
  \begin{align*}
    \check{\pi}'_y & {} \supseteq \{(4^m - 1) \cdot y + 2m + 10 \geq 0\} \tag{\eqref{eq:ub} with $j=2$ and $j'=0$} \label{eq:lb} \\
    \hat{\pi}'_y   & {} \supseteq \{(1 - 2^m) \cdot y - m + 10 \geq 0\} \tag{\eqref{eq:ub} with $j=0$ and $j'=1$}
  \end{align*}
  Thus, the resulting formula $\pi''$ contains the conjunct
  \begin{align*}
                 & (4^m-1) \cdot (10-m) - (1-2^m) \cdot (2m + 10) \geq 0 \\
    {} \equiv {} & -4^m \cdot m + 10 \cdot 4^m + 2^m \cdot 2m + 10 \cdot 2^m - m - 20 \geq
    0. \tag{\ensuremath{\lightning}}\label{eq:elim}
  \end{align*}
\end{example}

\subsubsection{Correctness of Variable Elimination:}\label{sect:Variable Elimination}
To see why this construction is correct, we show
that
\[
  \pi \equiv \pi' \; \text{for all large enough values of $m$},
\]
where $\pi' \Def (\pi \cup \pi_x) \setminus (\check{\pi}_x \cup \hat{\pi}_x)$.
More precisely, we prove that $\pi[m/c]$ is equivalent to $\pi'[m/c]$ for every $c\!\in\!\NN$ where $\sign(p[m/c])\!=\! \esign(p)$ for all coefficients $p$ in $\pi$.

Thus, if $m_0 \in \NN$ such that for all $m \geq m_0$ we have $\sign(p) = \esign(p)$, then $\pi$ is also equivalent to $\pi'$ for all $m \geq m_0$.
Hence, instead of checking unsatisfiability of $\pi$ for some sufficiently large value of $m$, we can instead check unsatisfiability of $\pi'$ for some sufficiently large value of $m$.

\paragraph*{Equivalence of $\pi[m/c]$ and $\pi'[m/c]$:}
To show that $\pi[m/c]$ is equivalent to $\pi'[m/c]$ if $\sign(p[m/c]) = \esign(p)$ for all coefficients $p$ in $\pi$, we can use the idea of Fourier-Motzkin elimination, since $\pi[m/c]$ is linear:

If all inequations are weak (i.e., built with $\geq$), then the inequations
\[
  p_1[m/c] \cdot x + t_1[m/c] \geq 0
\]
in $\check{\pi}_x[m/c]$ are equivalent to $x \geq \tfrac{- t_1[m/c]}{p_1[m/c]}$ (since $p_1[m/c] > 0$ by definition of $\check{\pi}_x$).
Similarly, the inequations
\[
  p_2[m/c] \cdot x + t_2[m/c] \geq 0
\]
in $\hat{\pi}_x[m/c]$ are equivalent to $x \leq \tfrac{- t_2[m/c]}{p_2[m/c]}$ (since $p_2[m/c] < 0$ by definition of $\hat{\pi}_x$).

Hence, $\pi[m/c]$ is equivalent to the conjunction of $(\pi \setminus (\check{\pi}_x \cup \hat{\pi}_x))[m/c]$ and the inequations
\begin{equation}
  \label{piPrime}
  \tag{\sc FM}
  \tfrac{- t_1[m/c]}{p_1[m/c]} \leq \tfrac{- t_2[m/c]}{p_2[m/c]}
\end{equation}
for all $p_1 \cdot x + t_1 \geq 0$ from $\check{\pi}_x$ and all $p_2 \cdot x + t_2 \geq 0$
from $\hat{\pi}_x$.\footnote{This does not hold for discrete sets like $\ZZ$, which is the
  reason why the approach of \Cref{sect:Variable Elimination} cannot be applied for loops over $\ZZ$.}

Note that if at least one of the two inequations $p_1 \cdot x + t_1 \geq 0$ or $p_2 \cdot x + t_2 \geq 0$ is strict, then \eqref{piPrime} is also strict.
Since $p_1[m/c] > 0$ and $p_2[m/c] < 0$, we get
\[
  \eqref{piPrime} \hspace{0.8em} \equiv \hspace{0.8em} -p_2[m/c] \cdot t_1[m/c] \geq -p_1[m/c] \cdot t_2[m/c] \hspace{0.8em} \equiv \hspace{0.8em} (p_1 \cdot t_2 - p_2 \cdot t_1 \geq 0) [m/c].
\]
Thus, $\pi[m/c]$ is equivalent to $(\pi \setminus (\check{\pi}_x \cup \hat{\pi}_x))[m/c] \cup \pi_x [m/c] = \pi'[m/c]$.

\subsection{Deciding Validity}
\label{sec:decide}

After eliminating all variables, we obtain a conjunction of inequations of the form $p > 0$ or $p \geq 0$ where $p$ is of the form \eqref{eq:coeff}.
Such a conjunction is unsatisfiable for large enough values of $m$ iff there is an inequation $p > 0$ or $p \geq 0$ where $\esign(p) = -1$ or an inequation $p > 0$ with $\esign(p) = 0$.
Thus, for the case where $\esign(p) = 0$, we have to distinguish between ``$>$'' and ``$\geq$''.

However, checking satisfiability in this way does not directly yield an instantiation of $m$ where $\pi$ is unsatisfiable.
Thus, we cannot directly obtain a corresponding constant bound $n_0 + \rb \cdot m$ on the runtime.
However, if $\pi$ is proven to be unsatisfiable, then to compute the constant bound on the runtime of the analyzed loop, the loop can simply be unrolled until its guard becomes unsatisfiable.
\begin{example}[Deciding Validity, \Cref{ex:elim} finished]
  For the formula \eqref{eq:elim}, we have $\esign(\ref{eq:elim}) = -1$, as $-4^m \cdot m$ is the dominant addend, since we
  have:
  \[
    \overbrace{(4,1)}^{-4^m \cdot m} >_{lex} \overbrace{(4,0)}^{10 \cdot 4^m} >_{lex} > \overbrace{(2,1)}^{2^m \cdot 2m} >_{lex} > \overbrace{(2,0)}^{10 \cdot 2^m} >_{lex} \overbrace{(1,1)}^{-m} >_{lex} \overbrace{(1,0)}^{-20}
  \]
  Thus, \eqref{eq:elim} is unsatisfiable for sufficiently large values of $m$, and hence \eqref{eq:no-conjunction1} is valid.
  Indeed, \eqref{eq:elim} is contradictory for all $m \geq 11$ (but not for $0 \leq m \leq 10$).
  Therefore, \Cref{ex:leading} has constant runtime, and we obtain the bound $n_0 + \rb \cdot m = 0 + 6 \cdot 11 = 66$, i.e., every run takes at most 66 iterations.
\end{example}
So in this section, we derived the following theorem.

\begin{theorem}[Constant Runtime over $\RR$]
  \label{thm:mon}
  Constant runtime of linear loops over $\RR$ with real eigenvalues is decidable.
\end{theorem}
Moreover, our approach can also be applied to loops over $\QQ$ without adaptions, as shown
by the following lemma.
\begin{lemma}[$\text{Constant over } \RR \iff \text{Constant over } \QQ$]
  \label{cor:Constant Runtime over R and Q}
  A linear loop with real eigenvalues has constant runtime over $\RR$ iff it has constant runtime over $\QQ$.
\end{lemma}
\begin{myproof}
  Constant runtime over the reals obviously implies constant runtime over the rationals.
  For the other direction, assume that the runtime over the rationals is bounded by the constant $c$.
  Then $\phi^{(0 \twodots c)} \Def \phi \land \up(\phi) \land \dots \land \up^c(\phi)$ is unsatisfiable over the rationals.
  Note that this formula only contains linear arithmetic.
  As formulas with linear arithmetic are satisfiable over the reals iff they are satisfiable over the rationals, $\phi^{(0 \twodots c)}$ is also unsatisfiable over the reals.
  Hence, the runtime over the reals is also bounded by the constant $c$.
\end{myproof}
\pagebreak[3]
Then with \Cref{thm:mon}, the following corollary is immediate.
\begin{corollary}[Constant Runtime over $\QQ$]
  Constant runtime of linear loops over $\QQ$ with real eigenvalues is decidable.
\end{corollary}

\section{Loops over $\ZZ$}
\label{sec:integers}

We now consider loops over the integers, i.e., where the coefficients in \eqref{loop:inhomogeneous} are from $\ZZ$, and where the variables range over $\ZZ$, too.
It is easy to see that all parts of the procedure presented in \Cref{sec:mon} immediately carry over to the integer setting except for one: We cannot eliminate variables via Fourier-Motzkin's technique as in \Cref{sec:elim}.
However, instead we can exploit decidability of ``one-parametric Presburger arithmetic'' \cite{bogart2017,mansutti25}.
This logic extends Presburger arithmetic with a single function $x \mapsto x \cdot m$, where $m$ is a free variable.
So the resulting formulas are non-linear polynomials w.r.t.\ $m$, but linear w.r.t.\ all other variables.
Thus, we can apply decidability of this logic to \eqref{eq:no-conjunction}, but not directly to \Cref{eq:constant}.

In our setting, recall that the bases $b_i$ of the exponential functions that occur in closed forms correspond to the eigenvalues of the loop.
Thus, if we restrict our attention to the eigenvalues $0$ and $1$, then all exponential functions vanish.
Moreover, if we also allow the eigenvalue $-1$, then after applying the simplification from \Cref{lem:simp}, the only remaining eigenvalues are again $0$ and $1$.
Thus, we obtain the following corollary.
\begin{corollary}[Constant Runtime over $\ZZ$]
  Constant runtime of linear loops over $\ZZ$ with eigenvalues from $\{-1,0,1\}$ is decidable.
\end{corollary}

\section{Related Work}
\label{sec:related_work}
Concerning related work on complete techniques, decidability of termination for linear single-path loops over $\RR$, $\QQ$, and $\ZZ$ was proven in \cite{Tiwari04}, \cite{Braverman06}, and \cite{Hosseini19}, respectively.
Furthermore, \cite{xuSymbolicTerminationAnalysis2013} and \cite{FMSD2023} presented decidability results on termination of solvable loops and of triangular weakly non-linear (twn) loops, respectively.
In contrast to the loops considered in our paper, such loops allow restricted forms of non-linearity.
Building on \cite{FMSD2023}, a technique to solve the (non-universal) halting problem (i.e., to decide termination for a \emph{fixed} input), and to infer polynomial runtime bounds for twn-loops was introduced in \cite{LPAR2020}.
The complete techniques for termination analysis and inference of upper bounds for twn-loops from \cite{LPAR2020,FMSD2023} were implemented in the tool \textsf{KoAT} \cite{lommen2022AutomaticComplexityAnalysis,lommen2023TargetingCompletenessUsing}.

Our work complements these results by providing a decision procedure for constant runtime.
Similar to \cite[Lemma 22]{lommen2023TargetingCompletenessUsing}, our approach can be extended to eigenvalues which are roots of reals (or of $\{-1,0,1\}$, in the integer case), i.e., to eigenvalues $\lambda$ where $\lambda^n\in\RR$ (or $\lambda^n\in\{-1,0,1\}$) for some $n\in\NN$.
In the future, it would be interesting to lift our results to update matrices with arbitrary eigenval\-ues (including complex ones that are no roots of reals).
However, in such a setting, several steps of our technique must be revised.
For example, the number of roots of poly-exponential expressions involving complex numbers cannot be bounded by a constant as in \Cref{sec:roots} and the quantifier elimination of \Cref{sec:qelim}\linebreak
to allow the subsequent application of Fourier-Motzkin elimination would no longer work.
Similarly, it would be interesting to generalize our results for $\ZZ$.

Note that the techniques mentioned above rely on the fact that the asymptotically dominant addend of closed-form expressions like \eqref{eq:polyexp} can easily be determined.
In \Cref{sec:esign}, we use a similar reasoning, but there, we are \emph{not} interested in the asymptotic behavior of \eqref{eq:polyexp}, but whether an inequation eventually gives rise to a lower or upper bound on a given variable $x$.
Thus, we are just interested in the dominant addend of $x$'s \emph{coefficient} when regarding \eqref{eq:polyexp} as a linear polynomial with poly-exponential coefficients.
Hence, our approach differs fundamentally from earlier techniques for deciding termination.
In particular and in contrast to earlier approaches (and as already mentioned in \Cref{sect:introduction}), we cannot exploit the ``eventual monotonicity'' of linear loops with real eigenvalues, as the initial number of ``non-monotonic'' steps (when ignoring the loop guard) is often \emph{not} bounded by a constant, even if the runtime of the loop is constant.
To see this, consider the following loop with constant runtime:
\[
  \wloop{0 \leq x \leq 10}{
    \mat{
      x \\
      y
    }
    \gets
    \mat{
      1 & 1 \\
      0 & 1
    }
    \mat{
      x \\y
    }
    +
    \mat{
      0 \\
      1
    }
  }
\]
Here, $x$ behaves monotonically as soon as $y$ becomes non-negative (as $y \geq 0$ implies $x \leq \up(x)$).
However, the number $k$ such that $\up^k(y) \geq 0$ holds is unbounded, as it depends on the initial value of $y$.

The complete technique for computing polynomial runtime bounds for twn-loops from \cite{LPAR2020}
allows for computing a polynomial upper bound for any terminating twn-loop.
However, these bounds are not necessarily tight, i.e., the resulting bound might be non-constant even if the runtime of the loop is constant.
Thus, our work is orthogonal to \cite{LPAR2020}.

There exist several techniques and tools for automatic complexity analysis of programs (with arbitrary multiple loops, if-then-else, non-determinism, \ldots) where the variables range over $\ZZ$.
Most of these tools, e.g., \textsf{CoFloCo} \cite{flores-montoya2016UpperLowerAmortized}, \textsf{KoAT} \cite{lommen2022AutomaticComplexityAnalysis}, \textsf{Loopus} \cite{sinn2017ComplexityResourceBound}, \textsf{MaxCore}
\cite{albert2019ResourceAnalysisDriven}, and \textsf{RAML}~\cite{RAML}, infer \emph{upper} bounds on the worst-case runtime complexity.
Thus, if such a tool computes a constant runtime bound, then the program is guaranteed to terminate within a constant number of steps.
In contrast, there also exist tools like \textsf{LOBER}
\cite{albert2021LowerBoundSynthesisUsing} or \textsf{LoAT} \cite{frohn2022ProvingNonTerminationLower}, which infer \emph{lower} bounds on the worst-case runtime complexity of loops or programs, respectively.
Consequently, such tools can \emph{dis}prove constant runtime.

\section{Evaluation and Conclusion}
\label{sec:conclusion}

\paragraph{Evaluation:}
To evaluate the practical applicability of our decision procedure, we implemented the approach from \Cref{sec:mon} in \textsf{Python}, using \textsf{SymPy} \cite{sympy} for all computations with algebraic numbers.
We did not yet implement the adaption to loops over $\ZZ$ from \Cref{sec:integers}, as we are not aware of any implementation of a decision procedure for one-parametric Presburger arithmetic.
For our evaluation, we used the benchmark set from \cite{frohn2020CalculusModularLoop} consisting of all 1495 linear single-path loops from the category \emph{Termination of Integer Transition Systems} of the \emph{Termination Problems Data Base} \cite{tpdb}, the benchmark suite used in the annual \emph{Termination and Complexity Competition (TermComp)} \cite{termcomp}.
After removing duplicates, we obtained 1336 loops.
All these loops have only real eigenvalues, which indicates that our approach is indeed applicable to a large class of examples.
1327 of them have eigenvalues from $\{-1,0,1\}$, which shows that even our seemingly restrictive result for loops over $\ZZ$ (see \Cref{sec:integers}) is very relevant from a practical point of view.

For our experiments, we compared the performance of our tool \textsf{Loopy} against all tools for complexity analysis of integer transition systems that participated in this year's \emph{TermComp}, i.e., \textsf{CoFloCo}, \textsf{KoAT}, and \textsf{LoAT}.
Note that these three tools consider initial values from $\ZZ$ instead of $\RR$ or $\QQ$.
Thus, if \textsf{LoAT} infers a non-constant \emph{lower} bound over the integers for a loop, then this loop also has non-constant runtime over the rationals and reals; conversely, whenever \textsf{Loopy} proves a constant bound over the rationals and reals, then this is also a constant \emph{upper} bound over the integers.
To the best of our knowledge, there are no other tools which analyze complexity of programs over $\RR$ or $\QQ$.
\textsf{Loopy} proved constant runtime for 76 loops and non-constant runtime for the 1260 remaining loops, with an average analysis time of 1.71 seconds on an AMD Ryzen 7 3700X octa-core CPU.
This shows that our decision procedure is not only of interest from a theoretical point of view, but it is also efficient on a standard benchmark set, which indicates its applicability in practice.
\textsf{CoFloCo} and \textsf{KoAT} were able to prove constant runtime for five and six of
these 76 loops, respectively, whereas \textsf{LoAT} proved non-constant runtime for 1226 loops over $\ZZ$.
All examples that were classified as non-constant by \textsf{LoAT} were also classified as non-constant by \textsf{Loopy}, i.e., there are no conflicting results.
Note that \textsf{CoFloCo} and \textsf{KoAT} do not implement dedicated techniques for proving constant bounds, and thus it is not surprising that they do not infer more constant runtime bounds.
Hence, our experiments demonstrate that tools for upper bounds can benefit from our decision procedure to improve their precision on such loops.
To download \textsf{Loopy} and for the detailed results of our evaluation, we refer to \cite{tool}.

\paragraph{Conclusion:}
We presented the first complete approach for the constant runtime problem of linear loops.
For loops over $\RR$ and $\QQ$, our decision procedure applies if the eigenvalues of the update matrix are real.
For loops over $\ZZ$, it covers eigenvalues from $\{-1,0,1\}$.
Our empirical evaluation shows the practicability of our approach.
In future work, we will try to add support for initial conditions for loops with real eigenvalues, which would yield a partial solution to the long-standing open problem whether the existence of multiphase-linear ranking functions is decidable for linear loops \cite{Ben-AmramGenaim19}.
Note that while termination of linear loops for fixed initial values corresponds to the positivity problem (which has been open for decades), the positivity problem is decidable for real eigenvalues \cite{MFCS2017}.
Moreover, we will try to extend our approach to arbitrary complex eigenvalues (without considering initial conditions).

\clearpage

\subsection*{Data Availability Statement}

An artifact including our implementation in the tool 
\textsf{Loopy} is available at \cite{tool}:
\[ \mbox{\url{https://doi.org/10.5281/zenodo.17313899}}
\]
This artifact allows to reproduce all 
results of our evaluation.

\bibliographystyle{splncs04}
\bibliography{refs}
\end{document}